# Isotope effects in supercooled $H_2O$ and $D_2O$ and a corresponding-states-like rescaling of the temperature and pressure

Greg A. Kimmel*

Physical Sciences Division, Pacific Northwest National Laboratory, P.O. Box 999, Richland WA, USA

99352

**Abstract**

Water shows anomalous properties that are enhanced upon supercooling. The unusual behavior is observed in both $H_2O$ and $D_2O$, however with different temperature dependences for the two isotopes. It is often noted that comparing the properties of the isotopes at two different temperatures (i.e., a temperature shift) approximately accounts for many of the observations – with a temperature shift of 7.2 K in the temperature of maximum density being the most well-known example. However, the physical justification for such a shift is unclear. Motivated by recent work demonstrating a "corresponding-states-like" rescaling for water properties in three classical water models that all exhibit a liquid-liquid transition and critical point (B. Uralcan, et al., *J. Chem. Phys*. **150**, 064503 (2019)), the applicability of this approach for reconciling the differences in temperature- and pressure-dependent thermodynamic properties of $H_2O$ and $D_2O$ is investigated here. Utilizing previously published data and equations-of-state for $H_2O$ and $D_2O$, we show that the available data and models for these isotopes are consistent with such a low temperature correspondence. These observations provide support for the hypothesis that a liquid-liquid critical point, which is predicted to occur at low temperatures and high pressures, is the origin of many of water's anomalies.

**Introduction**

Water is an unusual liquid that has been extensively investigated for over a century.[1, 2] Early work by Angell, Speedy and co-workers, which showed that many of water's anomalous properties are enhanced upon supercooling, perhaps signaling a singularity, has generated tremendous ongoing interest in this area.[3-11] The results of numerous experiments, theories, models, and simulations on supercooled water have been the subject of excellent reviews.[1, 10, 12-14] Currently, two related theories, the liquid-liquid critical point (LLCP) hypothesis and the singularity-free scenario, have the most experimental support.[1, 15-20] The LLCP hypothesis proposes that at low temperatures and high pressures water has two thermodynamically distinct (metastable) liquid phases, typically called the high- and low-density liquid or HDL and LDL, respectively, that are separated by a first order phase transition. The HDL-LDL coexistence line ends in a critical point – the LLCP. In that case, water beyond the critical point, is an inhomogeneous mixture of two locally-favored structures.[21, 22] For the singularity-free scenario, there are still two locally-favored structures that have different dependences on temperature and pressure, but these never lead to phase separation.

If water has an LLCP, then it belongs to the 3D Ising model universality class.[23-25] Several classical water models have been rigorously shown to have an LLCP.[14, 26, 27] Furthermore, various two-state models based on the physics associated with an LLCP can reproduce most of the available experimental data over a wide range of temperatures and pressures.[24, 25, 28-30] One of those models is also the basis for the recommended equation-of-state (EoS) for supercooled water by the International Association for the Properties of Water and Steam (IAPWS).[25]

Because the universal scaling associated with a critical point is in addition to a "normal" (non-diverging) component, experiments typically need to be done very close to the critical point to unambiguously observe the universal scaling behavior.[25] However, even far away from a critical point corresponding behavior is observed for many fluids (to a greater or lesser extent, depending on the fluid).[31-34] More generally, Pitzer[31] and Guggenheim[34] demonstrated the conditions necessary for "perfect" liquids to exhibit corresponding states and began the (ongoing) discussion of the deviations from this behavior expected for real liquids.[32, 33] In Pitzer's formulation, the Helmholtz free energy is a universal function, $F = F(T/A, V/R_0)$, where $A$ and $R_0$ are a characteristic energy and length scales associated with the molecular interaction potential, respectively. He noted that it is convenient, but not essential, to choose the liquid-vapor critical point (LVCP) temperatures, $T_c$, and volumes, $V_c$, as the scale parameters.

For water, the experimental data is relatively far from a putative LLCP, so assessing if it follows the scaling behavior for the 3D Ising model is challenging. Conversely, using the experimental observations to predict the location of any possible singularity is challenging – a point that was made even in the initial work of Angell and Speedy.[6] As discussed below, we are interested in the isotopes of water and how a possible second critical point influences their properties. It is important to note that we are not concerned with the universal power law scaling expected in the immediate vicinity of an LLCP. Instead, we are interested in investigating corresponding states (in Pitzer's sense) for the isotopes over a wider range of temperatures and pressures. It was noted early on that while various properties of $H_2O$ suggested a singularity at ~228 K, the corresponding results for $D_2O$ indicated a singularity at ~233 K.[10] Analysis of the melting curves of $H_2O$ and $D_2O$ led to similar conclusions.[35] Subsequent work suggested that shifting the temperature scale for $D_2O$ by the difference in temperature of maximum density, $\delta T_{MD} \cong 7.2$ K, between $D_2O$ and $H_2O$ (at atmospheric pressure) resulted in corresponding states for densities of the two isotopes.[36, 37] However, because a 7.2 K temperature shift was less successful for other properties, in practice $\delta T_{MD}$ came to be used as an adjustable parameter without any specific physical significance associated with it. Instead of shifting the temperatures to match the $T_{MD}$'s Limmer and Chandler suggested that the appropriate temperature and pressure scales for producing corresponding states in water and various classical water models were the $T_{MD}$ at atmospheric pressure and a reference pressure related to the enthalpy and volume changes for water upon melting (also at atmospheric pressure).[38]

Recently, Uralcan, et al., investigated possible scaling relationships between 3 classical water models that are known to have an LLCP (ST2, TIP4P/2005, and TIP5P).[39] By analyzing the patterns of extrema (density maximum and minimum, compressibility, etc.) in the *P-T* plane they found a "corresponding-states-like rescaling" for the pressure and temperature. Specifically, they found that for reduced temperatures, $\hat{T}$, and pressures, $\hat{P}$, the patterns of extrema for the models approximately collapsed onto universal curves when:

$$\hat{T} = \frac{(T - T_c)}{(T_{max} - T_c)} \quad (1a)$$

and

$$\hat{P} = \frac{(P - P(T_{max}))}{(P_c - P_{min})} \quad (1b).$$

In Eqn. 1, $T_c$ and $P_c$ are the critical temperature and pressure, respectively, for the LLCP of a given water model, while $T_{max}$, $P(T_{max})$ and $P_{min}$ are related to characteristics of the $T_{MD}$ line in the *P-T* plane for that model. Specifically, $P_{min}$ is the minimum pressure along the $T_{MD}$ line, and $T_{max}$ is maximum temperature on the $T_{MD}$ line, which occurs at $P = P(T_{max})$. Uralcan, et al., also included a small rotation in the *P-T* plane, which we will assume is small for the water isotopes and can be ignored. It is important to note that Eqn. 1 is different from the reduced temperatures and pressures associated with the liquid-vapor critical point: $\hat{T}_{LV} = \frac{T}{T_c^{LV}}$ and $\hat{P}_{LV} = \frac{P}{P_c^{LV}}$, where we have added the superscript *LV* to distinguish the LVCP from the LLCP. Because of these differences Uralcan et al., referred to Eqn. 1 as a "corresponding-states-like rescaling." However, we will simply refer to the low temperature "corresponding states" for H$_2$O and D$_2$O while keeping in mind this importance distinction.

Following the approach of Uralcan, et al.,[39] here we investigate whether a scaling relationship similar to Eqn. 1 produces low temperature corresponding states for the isotopes of water. If it does, the range of temperatures and pressures over which the correspondence holds between the isotopes will provide some evidence of the range over which a possible critical point exerts its influence on water's properties. Besides extensive data available on H$_2$O, considerable data is also available for D$_2$O, with considerably less data on other isotopes such as, H$_2^{18}$O, H$_2^{17}$O and D$_2^{17}$O. Therefore, we will consider the relationship between $H_2O$ and $D_2O$. To facilitate the analysis, we use published EoS's for supercooled $H_2O$[25] and supercooled $D_2O$.[40] We find that a simple scaling relationship for pressures and temperatures, which is analogous to Eqn. 1, produces corresponding states for $H_2O$ and $D_2O$ for pressures up to ~200 MPa and temperatures below ~300 K for various properties including the density, isothermal compressibility, and speed of sound. Furthermore, the resulting deviations from strict corresponding states follow patterns that are similar to the deviations observed for the corresponding states of $H_2O$ and $D_2O$ when they are referenced to the LVCP.

**Methods**

In Eqn. 1, there are 4 unknowns for each isotope, $T_c$, $T_{max}$, $P(T_{max})$, and $P_{min}$. Because these values are uncertain for $H_2O$ and $D_2O$ (assuming for now that the LLCP hypothesis is correct), the specific form of the reduced temperatures and pressures in Eqn. 1 was not convenient to use in the search for a correspondence between $H_2O$ and $D_2O$. Instead, it was convenient to work with the actual temperatures and pressures that were used as inputs to the EoS's for both $H_2O$ and $D_2O$. If Eqn. 1 describes low temperature corresponding states for $H_2O$ and $D_2O$, then there must be a linear relationship between the temperatures and pressures for the isotopes that produces the correspondence such that:

$$T_D = \beta T_H + \Delta T \qquad (2a)$$

$$P_D = \gamma P_H + \Delta P \qquad (2b)$$

where $T_i$ ($P_i$) for $i = H$ or $D$ refer to the temperatures (pressures) for $H_2O$ and $D_2O$, respectively. A second benefit of using Eqn. 2 to express the corresponding temperatures and pressures for $D_2O$ and $H_2O$ is that it is "agnostic" with respect to the possible existence and location of an LLCP.

As mentioned above, Eqn. 1 is different than the usual equations for $\hat{T}_{LV}$ and $\hat{P}_{LV}$. In the form of Eqn. 2, the corresponding temperatures and pressures for $D_2O$ and $H_2O$ relative to the LVCP, are

$$T'_D = \beta^{LV} T_H \qquad (3a)$$

$$P'_D = \gamma^{LV} P_H, \qquad (3b)$$

where $\beta^{LV} = \frac{T_C^{LV}(D_2O)}{T_C^{LV}(H_2O)}$ and $\gamma^{LV} = \frac{P_C^{LV}(D_2O)}{P_C^{LV}(H_2O)}$. Below, we will compare some aspects of the low temperature correspondence between $H_2O$ and $D_2O$ to the usual correspondence associated with the LVCP.

A given thermodynamic property, $X^i$, exhibits corresponding states if $X^H(T_H, P_H) = X^D(T_D, P_D)$, where $i = H$ or $D$ refers to $H_2O$ or $D_2O$, respectively. Because the thermodynamic response functions can be determined from the molar volume as a function of temperature and pressure, $V_m(T, P)$, we searched for suitable values for the parameters in Eqn. 2 – $\beta$, $\Delta T$, $\gamma$, and $\Delta P$ – that provided best match for $V_m^H(T_H, P_H) = V_m^D(T_D, P_D)$. To facilitate the search, it was important to use EoS's for $H_2O$ and $D_2O$ that included as much of the supercooled region as possible. For $D_2O$, we used the recent EoS developed by Hruby and co-workers that relied upon their high-quality measurements of the density and is valid from 254 K to 298 K and from atmospheric pressure to 100 MPa.[40] For the corresponding range of temperatures and pressures for $H_2O$, there are several choices for the EoS that give essentially identical molar volumes. We chose to use the EoS described in Holten, et al.,[25] which is the EoS for supercooled $H_2O$ recommended by the International Association for the Properties of Water and Steam (IAPWS). Below we will refer to these as the supercooled $H_2O$ or $D_2O$ EoS. For temperatures and pressures above the melting line of $H_2O$ and $D_2O$ (i.e., "normal" water), we used the REFPROP software package from the National Institute of Standards and Technology, which is based on the IAPWS EoS for $H_2O$ and $D_2O$, to calculate and compare the properties of interest.[41] We will refer to these as the NIST $H_2O$ and $D_2O$ EoS.

Because the densities are more commonly encountered than the molar volumes, below we compare the $D_2O$ densities – multiplied by the ratio of the molar masses – to the $H_2O$ densities. For a given $D_2O$ density, $\rho_i^D$, the mass scaled density is $\rho_i^{D\prime} = (m_{H2O}/m_{D2O}) \cdot \rho_i^D$, where $m_{H2O}$ and $m_{D2O}$ are the molar masses of $H_2O$ and $D_2O$. To optimize the parameters from Eqn. 2 (i.e., $\beta$, $\Delta T$, $\gamma$, and $\Delta P$), we calculated the $H_2O$ density at 1 K intervals from 249 to 293 K at 0.101325, 20, 40, 60, 80, and 100 MPa using the supercooled $H_2O$ EoS.[25] Those temperatures and pressures were then converted into their corresponding $D_2O$ values using Eqn. 2 for a trial set of parameters, and the corresponding $D_2O$ densities were calculated with the supercooled $D_2O$ EoS.[40] The parameters were then adjusted to minimize the average absolute deviation, $\Delta_{abs}$, between the $H_2O$ and $D_2O$ densities. For properties $X_i^H(T_H, P_H)$ and $X_i^D(T_D, P_D)$ calculated (or measured) at a series of points, $i$, $\Delta_{abs}$ was calculated as

$$\Delta_{abs} = (1/N_X) \sum_i^{N_X} \Delta_i^2 \qquad (4a)$$

$$\Delta_i = X_i^D / X_i^H - 1 \qquad (4b)$$

where $N_x$ is the number of data points and $\Delta_i$ is the relative deviation at each data point. Once the best fit values for the parameters were determined by comparing the densities, they were subsequently used without further adjustment to compare the isothermal compressibility, expansivity, speed of sound, and isobaric heat capacity of $H_2O$ and $D_2O$. We also extended the comparison outside the range of validity of the supercooled $D_2O$ EoS to investigate range of temperatures and pressures over which the low temperature correspondence provides reasonable estimates of the various properties.

In addition to comparing properties computed with the $H_2O$ and $D_2O$ EoS's, it was also useful to compare experimentally measured $H_2O$ properties at various temperatures and pressures, to the values calculated with the $D_2O$ EoS, at the corresponding $T_D$ and $P_D$. In some cases, we calculated the reduced residuals between the experimental data ($X_i^H(T_{H,i}, P_{H,i})$) and corresponding values calculated with the $D_2O$ EoS ($X_i^{D(EoS)}(T_{D,i}, P_{D,i})$) using published estimates of the absolute experimental uncertainty for the data. [30, 42] The reduced residual for a given data point $i$, $r_{X,i}$, is given by

$$r_{X,i} = \left[ X_i^H(T_{H,i}, P_{H,i}) - X_i^{D(EoS)}(T_{D,i}, P_{D,i}) \right] / \sigma_i \qquad (5)$$

where $\sigma_i$ is the associated absolute experimental uncertainty.[30] These values could then be compared to the reduced residuals calculated for the $H_2O$ data and $H_2O$ EoS.

**Results**

Figure 1 shows the correspondence between the $H_2O$ and (mass-scaled) $D_2O$ densities – $\rho^H = \rho^H(T_H, P_H)$ and $\rho^{D'} = \rho^{D'}(T_D, P_D)$, respectively – for the set of parameters that minimizes the average absolute deviation, $\Delta_{abs}$ (see Eqn. 4). The optimized parameters are: $\beta = 1.00576$, $\Delta T = 4.00$ K, $\gamma = 1.0187$, and $\Delta P = 10.362$ MPa. Fig. 1a shows $\rho^H$ calculated using (i) the supercooled H$_2$O EoS (solid red line) along with the NIST H$_2$O EoS (dashed red line).[25, 41] Similarly, $\rho^{D'}$ was calculated with EoS's for supercooled (open blue circles and diamonds) and normal D$_2$O (solid blue circles) states.[40, 41] Although the supercooled D$_2$O EoS is nominally valid for $P_D \leq 100$ MPa, the correspondence with the H$_2$O densities is also reasonably accurate up to 200 MPa and 300 K. Furthermore, the correspondence between normal H$_2$O and D$_2$O (i.e., above their melting points) calculated using the NIST EoS's is also generally good for T < ~300 K and P ≤ 200 MPa. Fig. S1 shows the relative deviation, $\Delta_i$, (see Eqn. 4b) between the densities calculate with the supercooled D$_2$O and H$_2$O EoS's. The differences are of the order of $10^{-4}$, and they show some systematic trends. For example, the differences between the densities for $P_D \leq 100$ MPa are generally the smallest for $T_H \sim 267$ K ($T_D \sim 273$ K). At lower temperatures, $\rho^{D'}$ is less than $\rho^H$ at low pressures, but larger at higher pressures, while the opposite trend is found at temperatures > 270 K. Given that both supercooled EoS's use polynomials in various ways, such systematic differences are not too surprising.

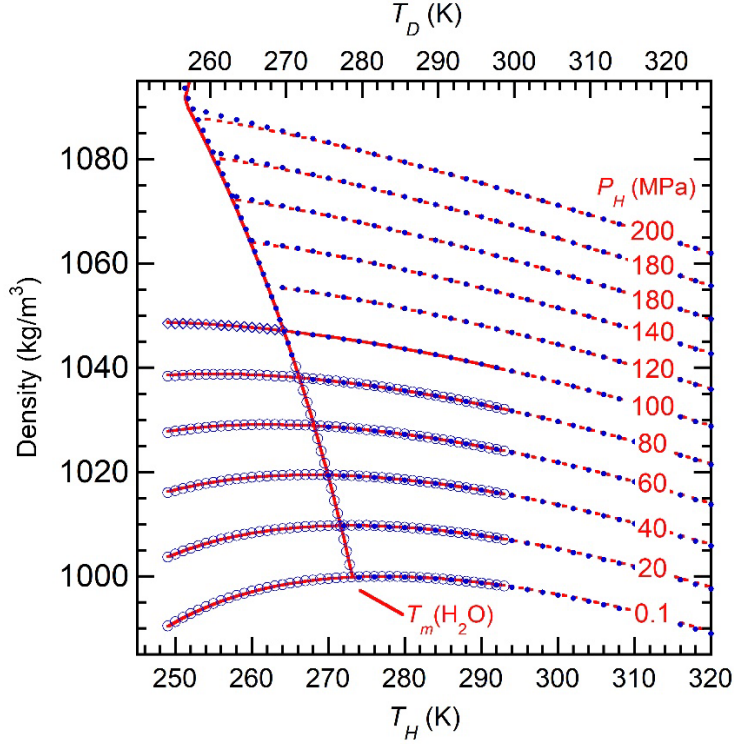

Fig. 1. Comparison of the $H_2O$ density, $\rho^H(T_H, P_H)$, to the mass-scaled $D_2O$ density, $\rho^{D'}(T_D, P_D)$. The bottom and top axes show the temperatures for $H_2O$ and $D_2O$, respectively. The $H_2O$ pressures, $P_H$, are shown in the figure, and the corresponding pressures for $D_2O$, $P_D$, are obtained from Eqn. 2b. The solid and dotted red lines show $\rho^H$ calculated with the supercooled $H_2O$ EoS of Holten, et al.,[25] and the NIST $H_2O$ EoS,[41] respectively. The open blue circles (diamonds) correspond to $D_2O$ densities calculated using the supercooled $D_2O$ EoS within (outside) its range of validity.[40] The filled blue circles show $D_2O$ densities calculate with NIST $D_2O$ EoS. The $H_2O$ and $D_2O$ densities along the $H_2O$ melting line, $T_m$, are also shown.

While the results in Fig. 1 compare densities calculated using the chosen $H_2O$ and $D_2O$ equations-of-state for supercooled water, it is also useful to compare the measured $H_2O$ densities to the corresponding $D_2O$ densities calculated using both the supercooled $D_2O$ EoS and the NIST $D_2O$ EoS. Caupin and Anisimov compiled experimental data for $H_2O$ densities along with estimates of the absolute experimental uncertainty, that they used to develop their EoS.[30] We used their results as input for the $D_2O$ EoS to calculate the corresponding $D_2O$ densities and the reduced residuals (see Eqn. 4). Figure 2 shows the results for the data of Hare and Sorensen,[43] and Sotani, et al.[44] For the range where the supercooled

D$_2$O EoS is valid, $-1 < r_{X,i} < 1$ for most of the data, with $r_{X,i}(\min) = -2.1$, and $r_{X,i}(\max) = 1.5$. The average absolute value of the reduced residuals, $ave(|r_{X,i}|)$, is 0.42. For comparison, using the supercooled H$_2$O EoS on the same data gives $ave(|r_{X,i}|) = 0.39$. As seen in the figure, including data with pressures up to 200 MPa (i.e., outside the valid range for the supercooled D$_2$O EoS), the correspondence is still quite good. It is interesting to note that, in contrast to the relative deviations between the supercooled H$_2$O and D$_2$O EoS's (Fig. S1), the reduced residuals calculated for supercooled D$_2$O EoS relative to the H$_2$O data do not show any obvious systematic trends (Fig. 2b).

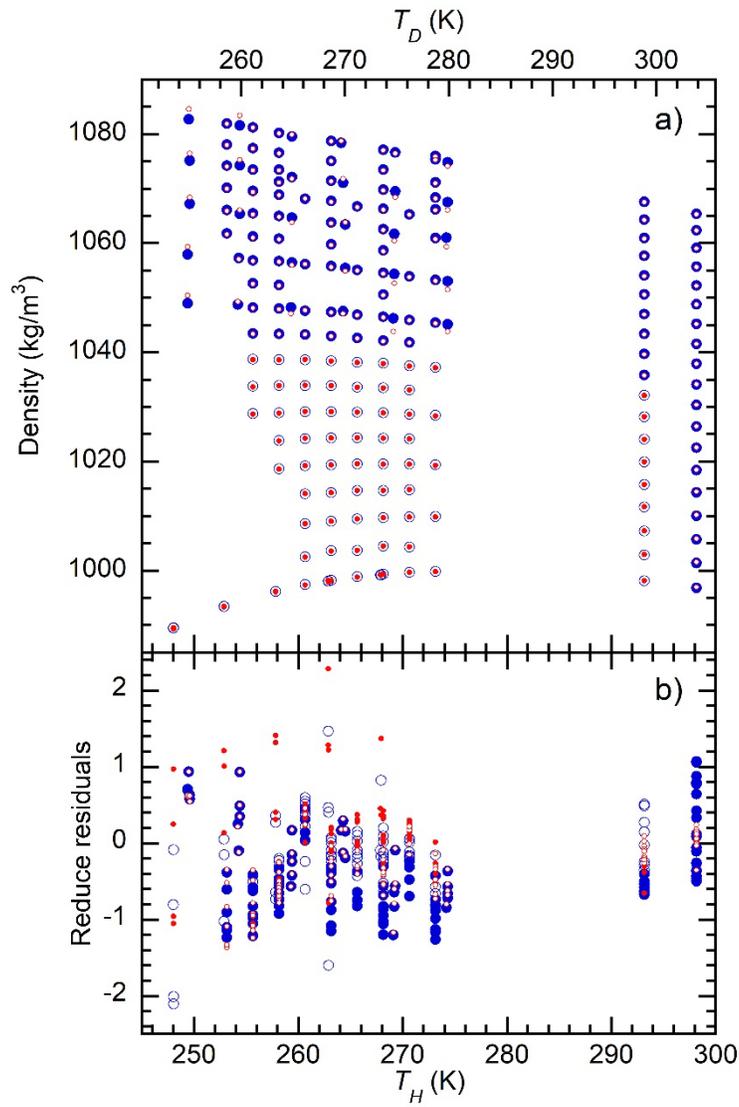

Fig. 2. a) Comparison of measured $H_2O$ densities (red circles),[43, 44] $\rho^H$, to the corresponding $D_2O$ densities, $\rho^{D'}$, calculated with the supercooled $D_2O$ EoS (blue circles).[25, 40] b) The reduced residuals (Eqn. 5) between the measured $H_2O$ and calculated $H_2O$ densities (red circles) are not appreciably different that residuals for the measured $H_2O$ densities and the calculated $D_2O$ values (blue circles). The open (filled) blue circles show $D_2O$ points that are within (outside) the range of validity of the supercooled $D_2O$ EoS, while the filled (open) red circles show points within (outside) the range of validity.

As discussed in the introduction, Uralcan, et al. used the lines of extrema in the $P$-$T$ plane, particularly focusing on the density maxima, to find approximate corresponding states for three classical water models.[39] Figure 3 shows the loci of the density maxima for $H_2O$, $L_{md}^H(T_H, P_H)$, (red circles) and the corresponding values for $D_2O$, $L_{md}^D(T_D, P_D)$, (blue diamonds).[40, 45, 46] The good overlap for the density maxima seen in Fig. 3 is unsurprising given the results shown in Fig. 1. However, based upon the results presented by Uralcan, et al., it also suggests that the other properties will show a similar correspondence.

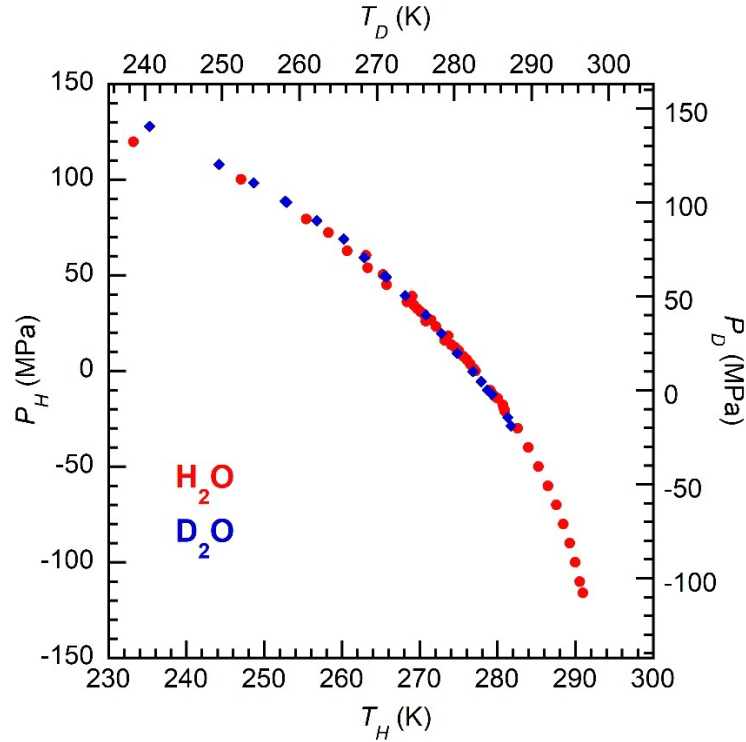

Fig. 3. Locus of maximum density for $H_2O$,[44, 46-50] $L_{md}^H(T_H, P_H)$, (red circles) and $D_2O$,[40, 49, 51] $L_{md}^D(T_D, P_D)$, (blue diamonds).

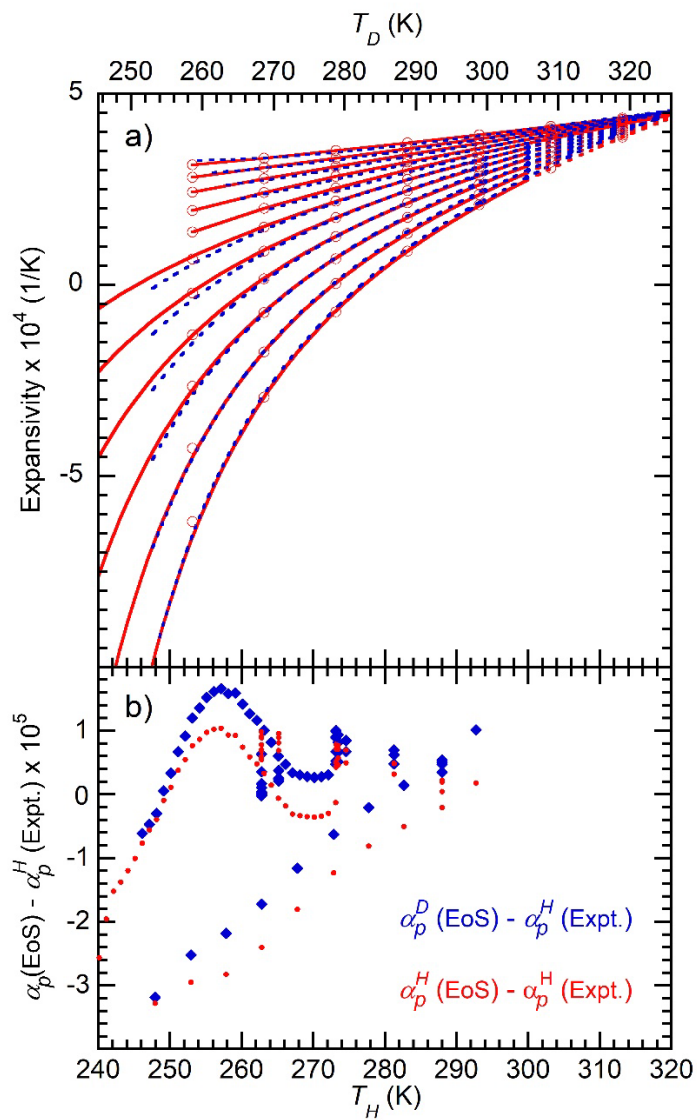

Fig. 4. a) Comparison of the expansivity calculated with $H_2O$ (red lines) and $D_2O$ (blue lines) EoS's. The supercooled EoS's were used for $T_H$, $T_D$ < 300 K and $P_H$, $P_D$ ≤ 100 MPa, otherwise the NIST EoS's were used. The open red circles show the expansivity for $H_2O$ derived from speed of sound measurements.[53, 54] b) Differences between $H_2O$ experimental expansivity data[43, 48, 52] and values calculated with the supercooled $D_2O$ (blue diamonds) and $H_2O$ (red circles) EoS's.

Generally, the various derivatives of the molar volumes with respect to temperature and pressure will be more sensitive to the deviations from the corresponding states picture, and thus could reveal more about the isotopic differences beyond what might be expected in a classical picture. Figure 4a compares the thermal expansivity, $\alpha_p = -\frac{1}{V}\left(\frac{\partial V}{\partial T}\right)_p$, for H$_2$O ($\alpha_p^H = \alpha_p^H(T_H, P_H)$) and D$_2$O ($\alpha_p^D = \alpha_p^D(T_D, P_D)$), calculated with the equations-of-state for both H$_2$O and D$_2$O. Fig. S2 shows a comparison of H$_2$O expansivity data with the supercooled D$_2$O EoS results, and Fig. 4b shows the differences between the experimental H$_2$O expansivity data[43, 48, 52] and the values calculated with the supercooled D$_2$O and H$_2$O EoS's (blue diamonds and red circles, respectively). The results in Fig. 4 indicate that the D$_2$O EoS's are largely able to reproduce the H$_2$O expansivity data at low temperatures. Furthermore, Fig. 4b suggests that the deviations of supercooled D$_2$O and H$_2$O EoS's with respect to the H$_2$O data are comparable.

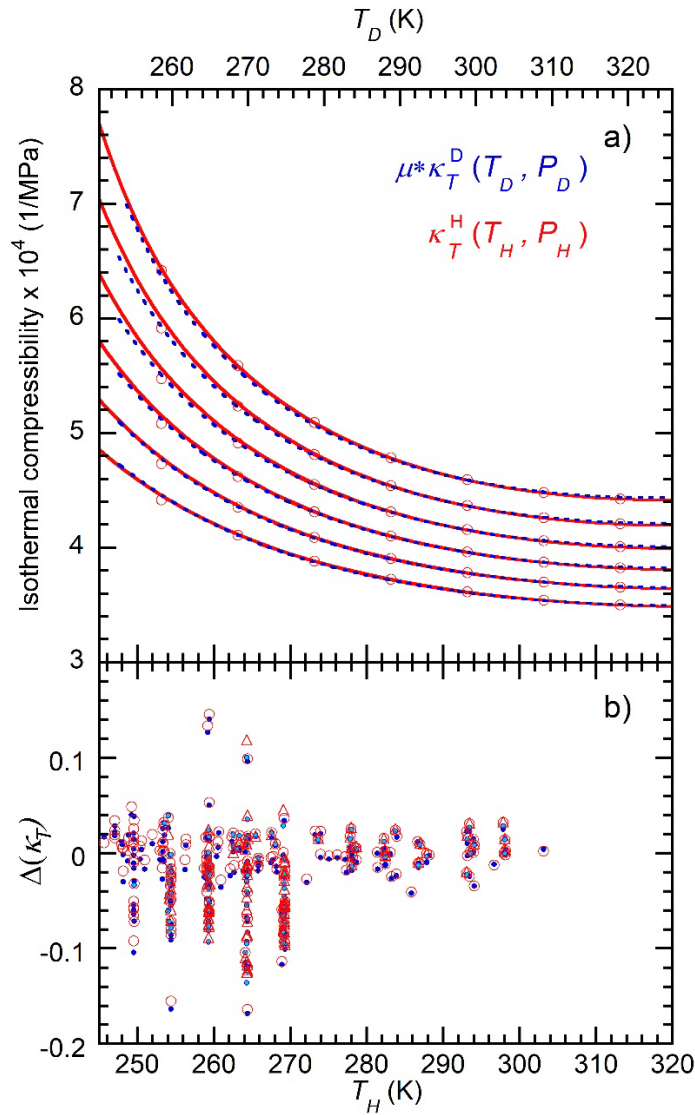

Fig. 5. a) Comparison of the isothermal compressibility calculated with the H$_2$O (red lines) and D$_2$O (blue lines) EoS's. The supercooled EoS's were used for $T_H$, $T_D$ < 300 K and $P_H$, $P_D$ ≤ 100 MPa, otherwise the NIST EoS's were used. The open red circles show the compressibility for H$_2$O derived from speed of sound measurements.[53, 54] b) Deviations between H$_2$O experimental compressibility data[6, 7, 50] and values calculated with the supercooled and NIST D$_2$O EoS's (dark and light blue circles, respectively) and supercooled and NIST H$_2$O EoS's (red circles and triangles, respectively) EoS's. The D$_2$O compressibility has been multiplied by $\mu$ = 1.015 in a) and for the calculation of the deviations in b).

Figure 5a compares the isothermal compressibility, $\kappa_T = -\frac{1}{V}\left(\frac{\partial V}{\partial P}\right)_T$, calculated with the supercooled and NIST EoS's for both isotopes. For D$_2$O, $\kappa_T^D$ is consistently less than the corresponding values for H$_2$O, but the trends versus temperature and pressure are nicely reproduced. As seen in Fig. 5a, an overall scale factor, $\mu \approx 1.015$, significantly improves the overlap (i.e., $\kappa_T^H \approx \mu \cdot \kappa_T^D$). Fig. S3 compares the H$_2$O compressibility data to the corresponding D$_2$O values calculated with the EoS's, and Fig. 5b shows the deviations of the H$_2$O and D$_2$O EoS's relative to the H$_2$O compressibility data (where the scale factor for $\kappa_T^D$, $\mu$, is included in the calculation).[6, 7, 50] As observed above for the density and the expansivity, the compressibility calculated using the D$_2$O EoS's using the low temperature correspondence produces similar deviations relative to the H$_2$O data compared to the H$_2$O EoS's, except in this case $\kappa_T^D$ is consistently about 1.5% smaller than $\kappa_T^H$ (see discussion below).

Accurate measurements of the speed of sound are available for H$_2$O, $w^H = w^H(T_H, P_H)$, over a wide range of temperatures and pressures (Fig. 6, red squares).[25, 42, 54] The speed of sound is inversely proportional to the square root of the density, so to compare between D$_2$O and H$_2$O, we use the mass-scaled D$_2$O speeds, $w^{D\prime} = \sqrt{m_{D2O}/m_{H2O}}\, w^D(T_D, P_D)$. However, after correcting for the mass differences, the corresponding states still show systematic differences in between the H$_2$O and D$_2$O. To illustrate this Fig. 6 shows $\lambda \cdot w^{D\prime}$ calculated with the supercooled D$_2$O EoS (blue circles). The value of the overall scale factor, $\lambda = 0.992$, which was determined by minimizing $\Delta_{abs}$ (see Eqn. 4) between $w^H$ and $\lambda \cdot w^{D\prime}$ over the range of validity for the supercooled D$_2$O EoS (shown by the black dotted lines in Fig. 6). For that range and with $\lambda = 0.992$, $\Delta_{abs}$ = 0.00093. The figure also shows $w^H$ (red lines) and $\lambda \cdot$ $w^{D\prime}$ (blue dashed lines) calculated with the NIST EoS for each isotope. At pressures above 100 MPa, where the supercooled D$_2$O EoS begins to deviate more noticeably, the correspondence for H$_2$O and D$_2$O calculated with the NIST EoS is still quite good.

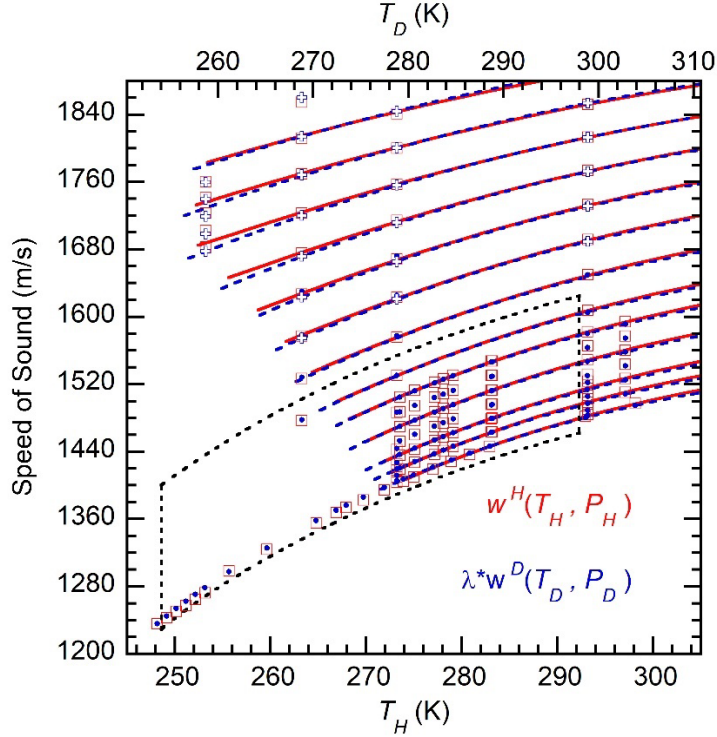

Fig. 6. Speed of sound comparison. H$_2$O data for Taschin, et al.,[55] Belogol'skii, et al.,[56] and Lin and Trusler (open squares circles)[53] and the corresponding values calculated with the supercooled D$_2$O EoS (blue circles) and the NIST D$_2$O EoS (blue crosses). The black dotted line shows the range where the supercooled D$_2$O EoS is valid. In addition to the expected correction by the square root of the masses, all the D$_2$O sound speeds are multiplied by $\lambda = 0.992$. This value minimizes the average absolute deviation, $\Delta_{abs}$, for the H$_2$O data compared to the supercooled D$_2$O EoS values in the range where it is valid. The red and dashed blue lines show the sound speeds calculated with the NIST EoS for H$_2$O and D$_2$O, respectively. For the NIST results, the H$_2$O pressures, $P_H$, are 0.101325, 10, 20, 40, 60, 75, 100, 125, 150, 175, 200, 225, and 250 MPa from bottom to top, and the corresponding D$_2$O pressures are calculated using Eqn. 2.

Figure 7 shows the isobaric heat capacities for H$_2$O ($c_p^H$) and D$_2$O ($c_p^D$) versus temperature for 0.1 MPa $\leq P_H \leq$ 400 MPa and the corresponding range of D$_2$O pressures. Red symbols show experimental results for H$_2$O,[30, 57-59] along with the results from the NIST H$_2$O EoS (solid red lines). (Other data, which extends to lower temperatures than the range of validity for the supercooled D$_2$O EoS, are not shown.)

For D$_2$O, $c_p^D$ was calculated with the supercooled EoS (solid blue line) for $P_D$ = 10.465 MPa (which corresponds to $P_H$ = 0.101325 MPa), while at higher pressures the NIST D$_2$O EoS was used (dashed blue lines). The agreement between the H$_2$O data and the corresponding values obtained with the D$_2$O EoS's is acceptable given that there is limited data and considerable uncertainty in the measurements at high pressures.[59] Furthermore, apparently the only data available for supercooled water is at atmospheric pressure.[40, 45] It is interesting to note that while the H$_2$O EoS's predict that $c_p^H$ decreases at low temperatures for $P_H$ > 100 MPa (see Fig. S4), the H$_2$O data and the D$_2$O EoS suggest that $c_p^H$ stops decreasing and perhaps goes through a minimum.[59] Because the heat capacity is likely to be sensitive to quantum effects,[60] experiments comparing supercooled H$_2$O and D$_2$O would be useful for developing a better understanding of how such effects influence the low temperature correspondence described here.

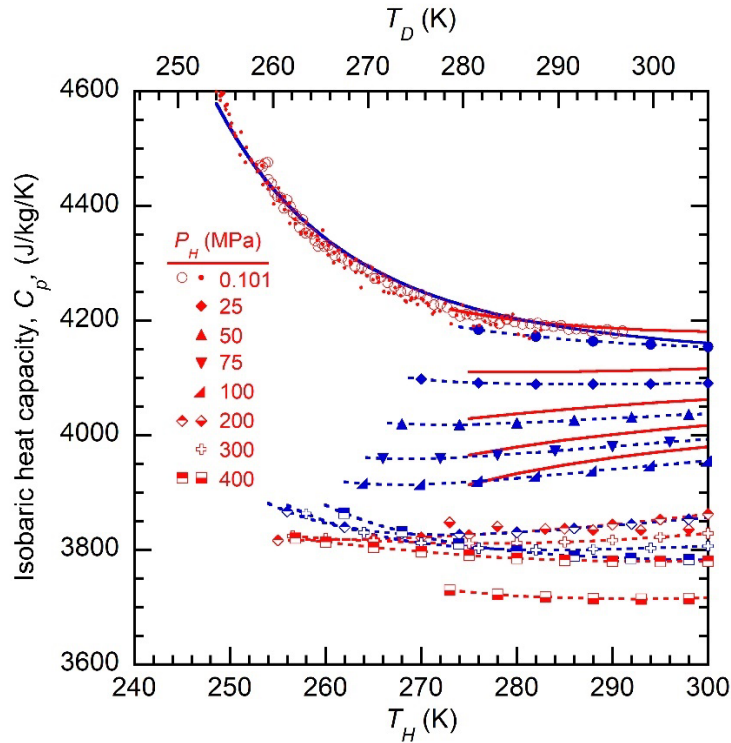

Fig. 7. Isobaric heat capacity for H$_2$O (red symbols and lines) and D$_2$O (blue symbols and lines). The solid red lines show $c_p^H$ at 0.101, 25, 50, 75, and 100 MPa (from the top down) calculate with NIST H$_2$O

EoS. The red symbols show experimental values for $c_p^H$.[57-59, 61] The solid and dashed blue lines show the corresponding values for $c_p^D$ from the supercooled and NIST D$_2$O EoS's, respectively.

The low temperature correspondence between H$_2$O and D$_2$O shows systematic deviations for some properties, such as the compressibility (Fig. 5) and the speed of sound (Fig. 6). However, it is noteworthy that using the standard corresponding states scaling associated with the LVCP (see Eqn. 3) also results in systematic deviations between H$_2$O and D$_2$O for these properties. For example, for temperatures near the LVCP and H$_2$O pressures ≤ 100 MPa, the speed of sound for D$_2$O is systematically less than H$_2$O such that an overall scale factor of ~1.015 significantly reduces the differences (Fig. S5). This is compared to the results for low temperatures, where, as discussed above, a scale factor of 0.992 produces a better correspondence (Fig. 6). Similarly, multiplying the D$_2$O compressibility, $\kappa_T^D$, by 0.982 reduces $\Delta_{abs}$ in the vicinity of the LVCP (see Fig. S6), compared to a scale factor of ~1.015 for the low temperature correspondence (see Fig. 5). Because the isothermal compressibility is proportional to the square of the volume fluctuations,[12] the experimental results show that the fluctuations for D$_2$O are smaller (larger) than the corresponding fluctuations for H$_2$O near the LLCP (LVCP). On the other hand, the expansivity is proportional to the product of the volume and entropy fluctuations,[12] so the apparent lack of systematic differences between $\alpha_p^H$ and $\alpha_p^D$ (Fig. 4) indicates that the reduced volume fluctuations in D$_2$O are compensated by entropy fluctuations. While the heat capacity is proportional to the square of the entropy fluctuations, the large uncertainties in both the data and the EoS predictions for $c_p^H$ and $c_p^D$ make it difficult to assess their relative magnitudes in supercooled water.

Above ~300 K, the low temperature correspondence gets progressively worse (as expected). Conversely, the correspondence predicted between H$_2$O and D$_2$O near the LVCP gets worse at lower temperature. Therefore, it is instructive to consider the temperatures at which the low and high temperature correspondences produce comparable results. Figure 8 shows the differences in densities between D$_2$O and H$_2$O – calculated with the NIST EoS's – using the low temperature correspondence (Eqn. 2), $\delta\rho(LL) = \rho^{D'} - \rho^H$ (dark blue symbols), and the liquid-vapor correspondence (Eqn. 3), $\delta\rho(LV) = \rho^{D'} - \rho^{H}$ (light blue symbols). For the range of pressures shown, the low temperature correspondence is more accurate for $T_H <$ 347 K, while the liquid-vapor correspondence is more accurate for $T_H >$ 378 K. The red circles in Fig. 8 show the points at which deviations calculated using the low and high temperature correspondences cross. It is interesting to note that at ambient pressure, this temperature is ~350 K, which is near the isothermal compressibility minimum for $H_2O$. The isothermal compressibility minimum has been suggested to be an indicator of the point at which the two-state character of water begins to have appreciable influence on the properties of water. (However, see the discussion below regarding the transition between "two-state" and "one-state" descriptions of liquid water.)

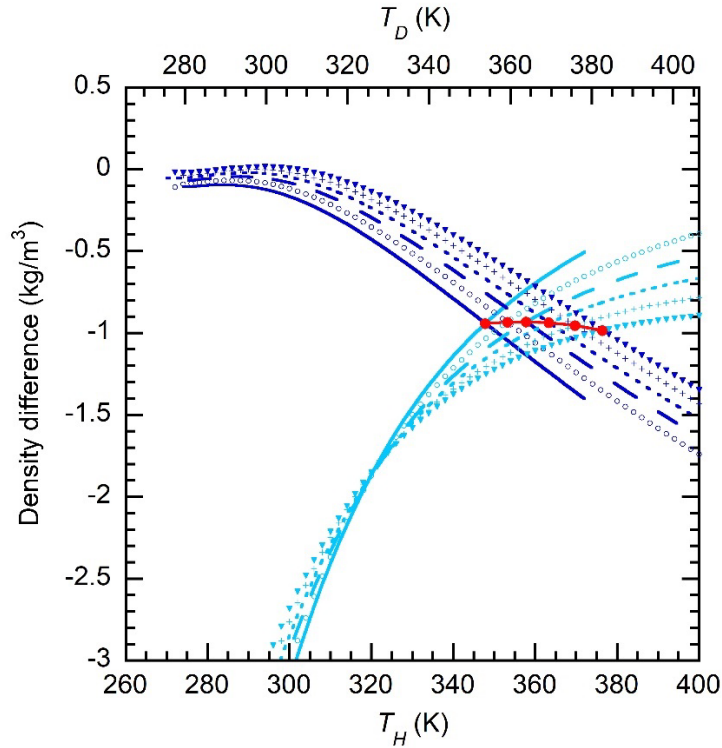

Fig. 8. Density differences between $H_2O$ and $D_2O$ calculated using the low temperature correspondence (Eqn. 2, dark blue symbols and lines) and the liquid-vapor correspondence (Eqn. 3, light blue symbols and lines). The results shown are for $P_H =$ 0.101325 MPa (solid lines), 20 MPa (circles), 40 MPa (dashed lines), 60 MPa (dotted lines), 80 MPa (+'s), and 100 MPa (diamonds). The low temperature correspondence is more accurate for $T_H <$ ~350 K. The red circles show where the low and high temperature correspondence cross at each pressure.

Previous investigations of the properties of stretched water have noted the influence of the liquid-vapor spinodal on water's thermodynamic properties.[9, 29, 30, 62-64] For example, Uralcan, et al., found a correlation between the liquid-vapor spinodal and the LLCP in three classical water models.[39] In a two-state model, the liquid-vapor spinodal of the high-temperature state contributes a term to its Gibb's free energy, which then influences the equilibrium fraction of each state as a function of temperature and pressure.[29, 30, 64] The low temperature correspondence between $H_2O$ and $D_2O$ also suggests a connection between the two critical points. Figure 9 shows several lines of extrema for $H_2O$ and $D_2O$ versus reduced temperature, $\hat{T}$, and pressure, $\hat{P}$. For this figure, Eqn. 1 has been used to calculate $\hat{T}$ and $\hat{P}$, and the values of $T_c$, $T_{max}$, $P_c$, $P(T_{max})$, and $P_{min}$ for $H_2O$ were taken from Table III and Fig. 13 in ref.[30] The values for $D_2O$ in Eqn. 1 were then calculated from the $H_2O$ values using the low temperature correspondence (Eqn. 2). The red/blue diamond shows the location of the LLCP for $H_2O$ and $D_2O$ (which are the same, by construction), while the red and blue squares show the LVCP or $H_2O$ and $D_2O$, respectively. It is noteworthy that using the low temperature correspondence places the $D_2O$ LVCP nearly on $H_2O$ liquid-vapor spinodal and suggests the liquid-vapor spinodal for $D_2O$ will closely follow the $H_2O$ spinodal. This observation is similar to the correlation between distances from the LLCP to various points on the liquid-vapor spinodal for three water models found by Uralcan, et al.[39]

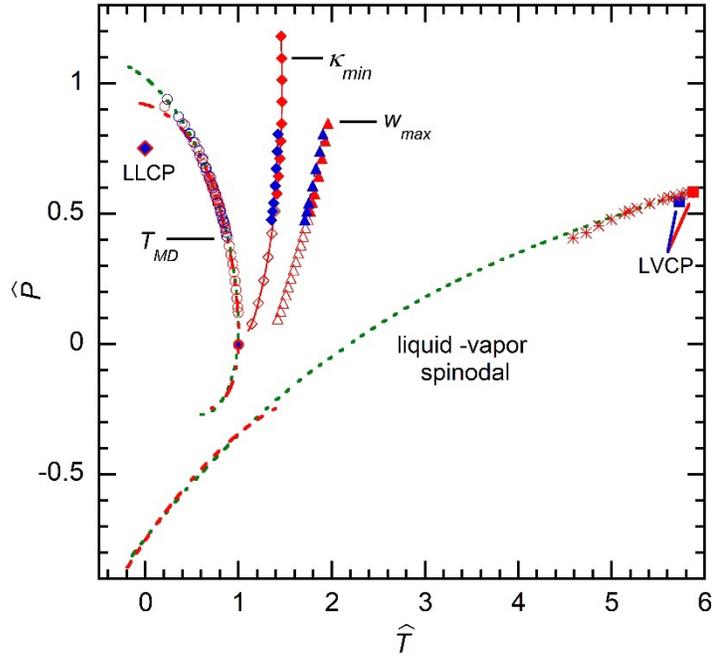

Fig. 9. Extrema lines for thermodynamic properties of $H_2O$ and $D_2O$ versus reduced temperature, $\hat{T}$ and pressure, $\hat{P}$ (see Eqn. 1). The open red and blue circles show the loci of density maxima, $L_{md}^H$ and $L_{md}^D$, for $H_2O$ and $D_2O$, respectively.[40, 46, 47, 50] The red stars show the location of $H_2O$ liquid-vapor spinodal[9] near the $H_2O$ liquid-vapor critical point (red square). In lieu of reliable data for the liquid-vapor spinodal at low temperatures, the dotted red and green lines shows the spinodal calculated for the TIP4P/2005 model[30] and derived from a two-state model,[64] respectively. The NIST EoS's were used to find the compressibility minima and speed of sound maxima at positive pressures for $D_2O$ (blue diamonds and triangles) and $H_2O$ (red diamonds and triangles). For negative pressures, the data of Pallares, et al. are shown as open red diamonds and triangles for the $\kappa_{min}$ and $w_{max}$, respectively.[46]

**Discussion**

If water has an LLCP, then both $H_2O$ and $D_2O$ should exhibit the universal scaling expected of the 3D Ising model in the immediate vicinity of the critical point. For example, Holten, et al.,[25] showed that a two-state model based on a LLCP could account for the experimental data for both $H_2O$ and $D_2O$. They also noted that "(w)hile the critical part of the thermodynamic properties of $H_2O$ and $D_2O$ follow the law of corresponding states (the critical amplitudes *a* and *k* are the same) the regular parts do not follow this law." The results presented should be consistent with those observations. In particular, a consistent treatment of the "regular parts" could presumably be developed that also accounts for the low temperature correspondence discussed here and thus account for the behavior in the immediate vicinity of the critical point and over the larger range of temperatures and pressures. Further research is needed to explore this possibility in detail.

For the analysis presented here, the parameters in Eqn. 2 for the corresponding states were determined by minimizing the difference in the molar volumes using the supercooled $H_2O$ and $D_2O$ EoS's.[25, 40] Including other properties and data in the optimization, using a different choice for weighting the contribution of various data, and/or changing the range of temperatures and pressures would undoubtedly change the specific values obtained for $\beta$, $\gamma$, $\Delta T$, and $\Delta P$. One possible outcome of such changes could be a reduction in some of the systematic differences observed between the isotopes for properties such as the speed of sound and the isothermal compressibility that were described above. However, while further refinements of the correspondence described here will be valuable, they seem unlikely to change the main observation, which is that the properties of $H_2O$ and $D_2O$ at supercooled temperatures are brought into correspondence with a linear scaling of temperatures and pressures that includes a non-zero offset term for each (see Eqn. 2).

The differences between $H_2O$ and $D_2O$ are ultimately derived from nuclear quantum effects (NQEs).[65-68] For example, classical simulations cannot predict the changes in $T_{MD}$ for the different isotopes of water.[65] Recent simulations that include NQEs on the thermodynamic properties of $H_2O$ and $D_2O$ for a wide range of pressures and temperatures (including the supercooled states) largely reproduce the experimental results for the density and isothermal compressibility.[68] In addition, those calculations follow the low temperature correspondence described here reasonably well. In particular, the $T_{MD}$ and liquid-vapor spinodal lines, which are determined from the simulations, essentially overlap (see Fig. S7). (The corresponding locations of the LLCP for $H_2O$ and $D_2O$, which are determined by fitting the simulation results to a two-state model, are also similar, but the agreement is not as good.) Because the results presented here are based on data at $T > 235$ K (at 0.1 MPa), it leaves open the possibility that the low temperature correspondence found for $H_2O$ and $D_2O$ might break down at even lower temperatures.

For example, previous results found that, while NQE's are important in the description of low-density amorphous ice (LDA) and hexagonal ice at very low temperatures, the difference between quantum and classical MD simulations were less important at higher temperatures.[69]

Because the potential energy surface (PES) for a collection of water molecules does not depend on the isotope, the differences between the isotopes comes from their behavior on the PES.[70] In this context, it is useful to consider supercooled water's properties in the potential energy landscape (PEL) framework.[71] The PEL is a hypersurface that represents the potential energy for a system as a function of the coordinates of all the atoms in the system. At sufficiently low temperatures, liquids primarily reside in local minima on the PEL, and their behavior is dominated the properties of these minima and the infrequent transitions it makes between minima. These properties (such as the number of minima versus energy and their curvature) can be used to determine the partition function for the liquid. For supercooled water, a simple model for the PEL (the Gaussian PEL) can account for water's anomalous properties and is consistent with the results of classical MD simulations and two-state models of the LLCP.[72, 73] In the PEL framework, isotopes of water will show corresponding behavior if they inhabit portions of the PEL that are similar (statistically). The low temperature correspondence describe here indicates that this occurs when $D_2O$ is at slightly higher pressures and temperatures relative to $H_2O$. These differences can presumably be modeled by differences in the zero-point energies associated with the local minima in the PEL and also anharmonic effects on the vibrational component of the free energy.[71, 73] The low temperature correspondence between the isotopes is similar to the widely noted idea that the structure of liquid $D_2O$ at a given temperature (above the melting point) is similar to that of $H_2O$ at a somewhat higher temperature. The primary difference between the low and high temperature cases is that as the temperature increases, the influence of the local minima in the PEL on the thermodynamics (and dynamics) is reduced, the fraction of the low-temperature structural motif decreases, and temperature-dependent changes in structure of the (essentially single-component) high temperature liquid can account for the isotopic differences.[66] Of course defining the transition when water is best described as an inhomogeneously broadened, single-component liquid and one that is best described by a two-state model depends on one's definitions and is subject to considerable debate.[2, 74, 75]

The results presented here indicate that there is an approximate, low temperature correspondence for the thermodynamic properties of $H_2O$ and $D_2O$. It also well-known that many of water's dynamic properties are potentially consistent with the LLCP hypothesis,[1, 6] with the $D_2O$ results typically showing a shift to higher temperatures (for the same pressure) that is similar to those observed for the thermodynamic properties.[10, 76-78] The amount of highly accurate, pressure-dependent dynamic data that is available for both $H_2O$ and $D_2O$ limits the ability to perform a detailed comparison of the low temperature correspondence in most cases. However, for the self-diffusion in supercooled $H_2O$ and $D_2O$,[76, 79] the low temperature correspondence appears to provide a reasonable description of the results (see Fig. S8). More work is needed to assess the extent to which the low temperature correspondence applies to other dynamical properties.

**Conclusions**

When comparing thermodynamic properties of supercooled $H_2O$ and $D_2O$, a simple linear relationship between the temperatures and pressures of the isotopes (Eqn. 2) produces a correspondence such that $X^H(T_H, P_H) \approx X^D(T_D, P_D)$, where $X^H$ and $X^D$ are properties, such as the molar volume, expansivity, isothermal compressibility, and speed of sound, for $H_2O$ and $D_2O$. This approximate, low temperature correspondence for the isotopes, which is distinct from the usual corresponding states associated with the liquid-vapor critical point, is generally good for temperatures below ~300 K and pressures below ~200 MPa. The most plausible physical origin for the low temperature correspondence is a liquid-liquid critical point for supercooled water. Based on the range of temperatures and pressures that produce a correspondence between properties of $H_2O$ and $D_2O$, these results support the idea that some of water's most notable anomalies, such as the existence of the density maximum at near ambient temperatures, are related to the LLCP in the deeply supercooled region.

**Author Declarations**

**Conflict of Interest**

The author has no conflict of interests to disclose.

**Acknowledgement**

The author would like to thank Bruce D. Kay, Nicole R. Kimmel, and Frédéric Caupin for helpful discussions, and Jan Hruby for help implementing his $D_2O$ equation-of-state. This work was supported by the U.S. Department of Energy (DOE), Office of Science, Office of Basic Energy Sciences, Division of Chemical Sciences, Geosciences, and Biosciences, Condensed Phase and Interfacial Molecular Science program, FWP 16248.

**Data Availability**

No new data is reported in this work.

**ORCID**

Greg A. Kimmel: 0000-0003-4447-2400

**Supplemental Information for:**

Isotope effects in supercooled H₂O and D₂O and a corresponding-states-like rescaling of the temperature and pressure

Greg A. Kimmel*

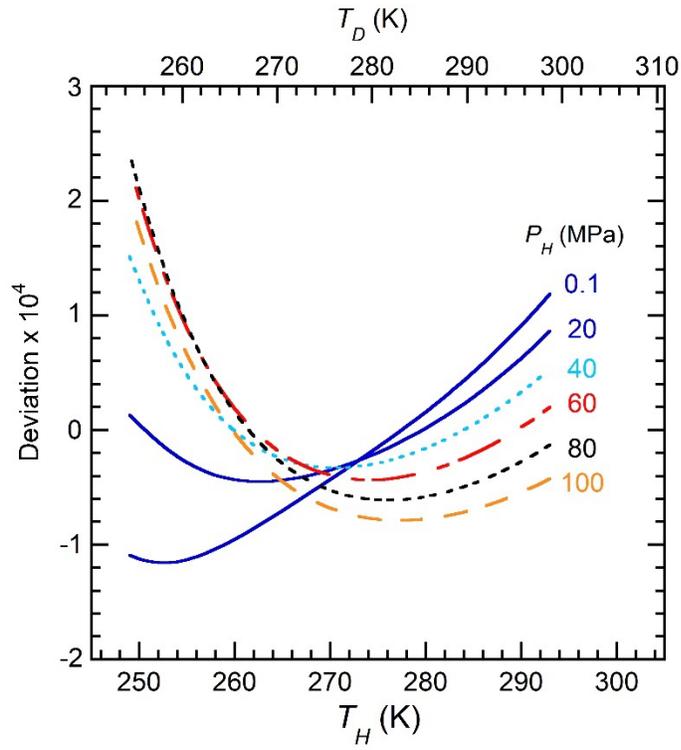

Fig. S1. Deviations (see Eqn. 4b in the main text) between the supercooled H₂O and D₂O equations of state.

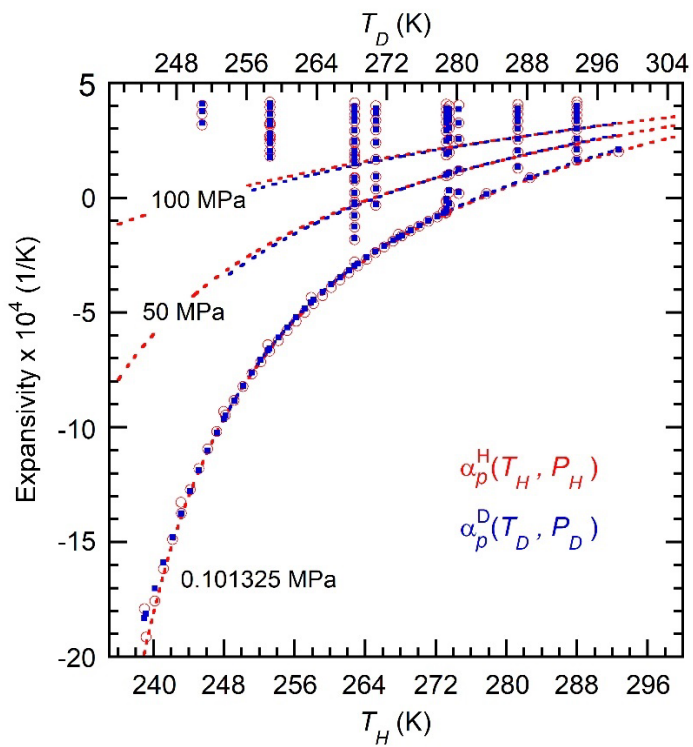

Fig. S2. $H_2O$ expansivity data (red circles) and the corresponding values calculated with the supercooled $D_2O$ EoS (blue squares). Fig. 4b show the differences between these values for the range of validity of the supercooled $D_2O$ EoS. The red and blue dashed lines show the expansivity at $P_H$ = 0.101325, 50 and 100 MPa (and the corresponding $D_2O$ pressures) calculated with supercooled $H_2O$ and $D_2O$ EoS, respectively.

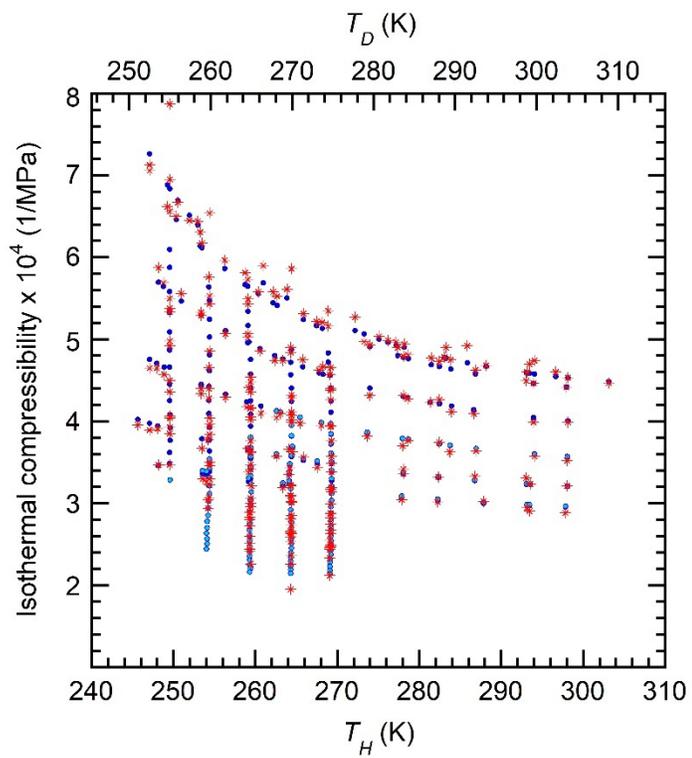

Fig. S3. H$_2$O compressibility data[1-3] (red stars) and the corresponding values calculated with the supercooled D$_2$O and NIST EoS (dark and light blue circles, respectively).

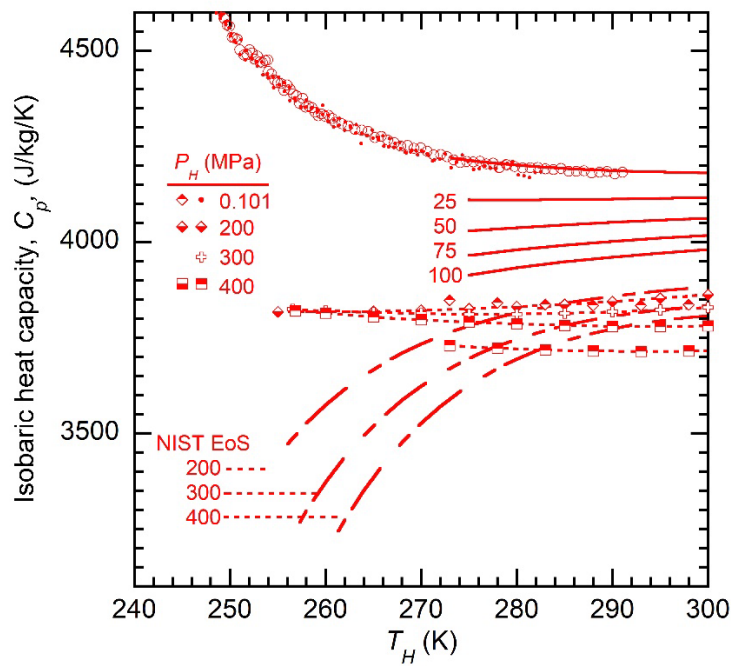

Fig. S4. Isobaric heat capacity for $H_2O$ comparison of experimental values to NIST $H_2O$ EoS at 200, 300, and 400 MPa. At these higher pressures, the NIST EoS decreases as the temperature decreases whereas the $H_2O$ data is approximately independent of temperature.

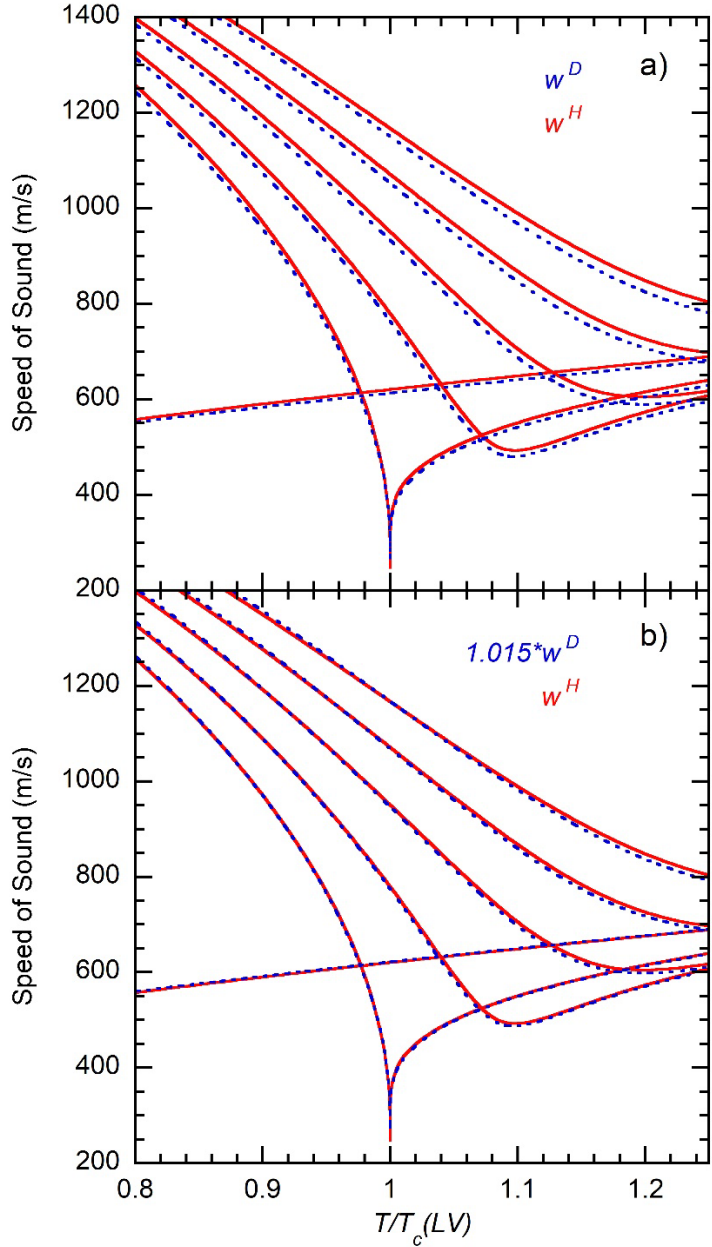

Fig. S5. Speed of sound, $w$, for $H_2O$ (red lines) and $D_2O$ (blue lines) versus the reduced temperature, $\hat{T}_{LV} = \frac{T}{T_c^{LV}}$, for reduced pressures, $\hat{P}_{LV} = \frac{P}{P_c^{LV}}$, of 0.0046, 1.0, 1.81, 2.72, 3.63 and 4.53, where the critical pressures for $H_2O$ and $D_2O$ are 22.065 and 21.671 MPa, respectively. a) To account for the expected mass effects, the speed of sound for $D_2O$ has been multiplied by square root of the masses. However, the mass-scaled $D_2O$ speeds are consistently less than the corresponding $H_2O$ values. b) Including an additional scaling factor, here taken to be 1.015, significantly improves the correspondence between the isotopes for the range of temperature and pressures shown here. Note that for the low temperature correspondence (Fig. 6), the scaling factor, $\lambda$, is less than 1.

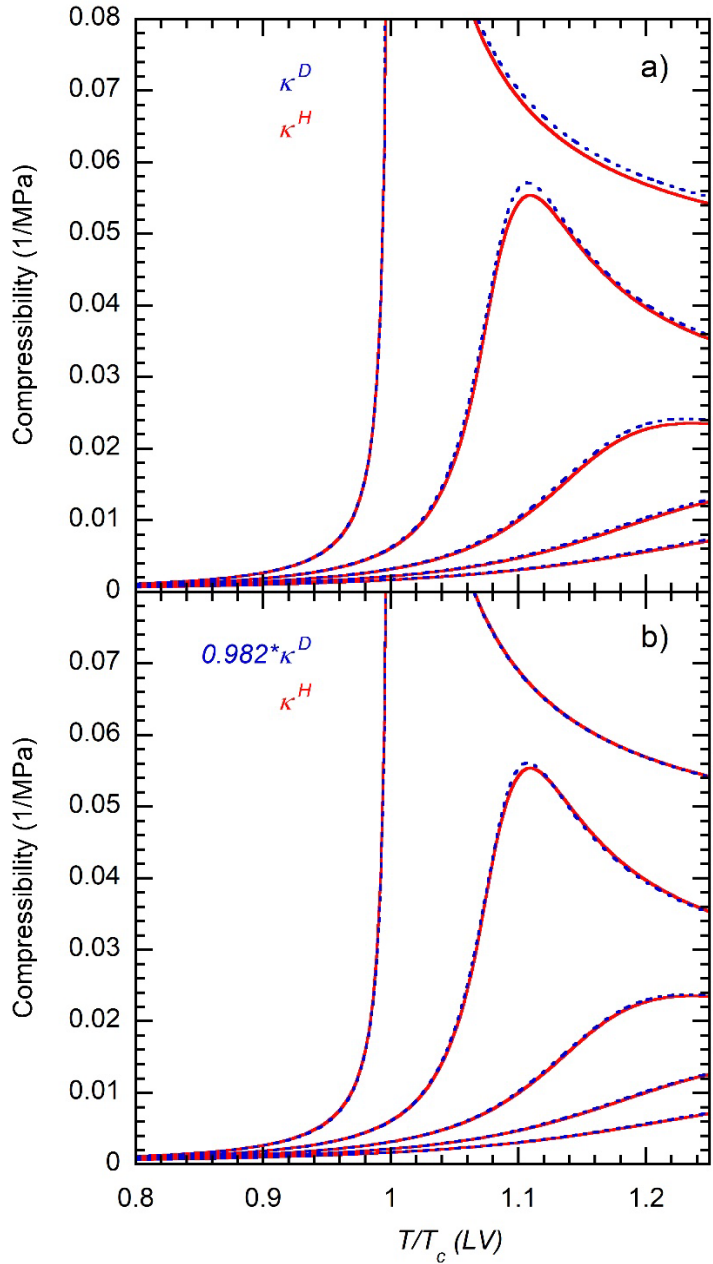

Fig. S6. Compressibility for $H_2O$ (red lines) and $D_2O$ (blue lines) versus the reduced temperature, $\hat{T}_{LV} = \frac{T}{T_c^{LV}}$, for reduced pressures, $\hat{P}_{LV} = \frac{P}{P_c^{LV}}$, of 0.0046, 1.0, 1.81, 2.72, 3.63 and 4.53, where the critical pressures for $H_2O$ and $D_2O$ are 22.065 and 21.671 MPa, respectively. a) The $D_2O$ compressibilities are consistently larger than the corresponding $H_2O$ values. b) Including an additional scaling factor, here taken to be 0.982, significantly improves the correspondence between the isotopes for the range of temperature and pressures shown here. Note that for the low temperature correspondence (Fig. 5b), the scaling factor is greater than 1.

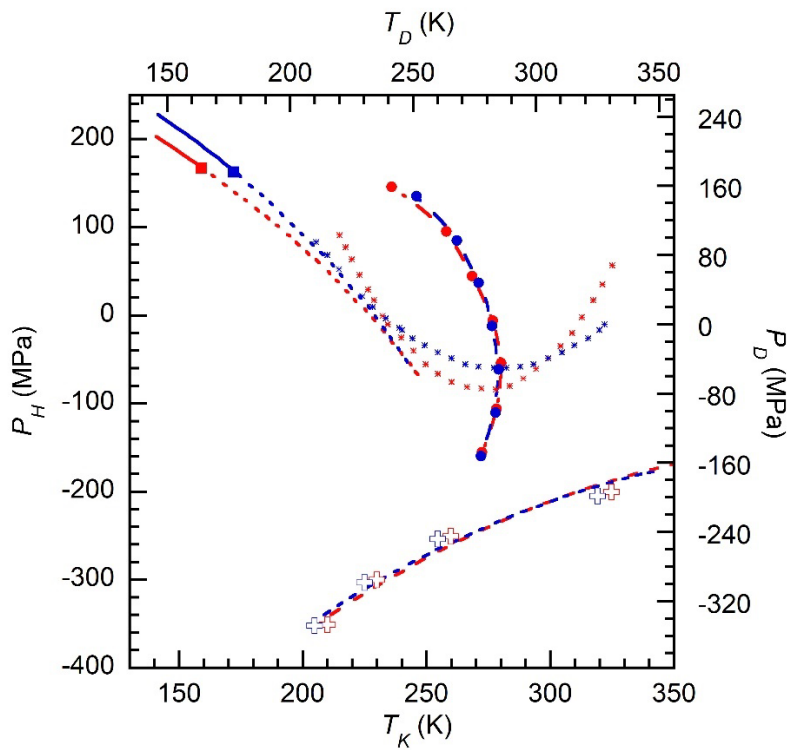

Fig. S7. Path integral molecular dynamics (PIMD) simulations of thermodynamic properties for $H_2O$ (red symbols and lines) and $D_2O$ (blue symbols and lines) for the q-TIP4P/F water model by Eltareb, Lopez, and Giovambattista.[4] The circles show the calculated $T_{MD}$ points (with lines to guide the eye). The crosses show the points where spontaneous cavitation occurred in the simulations, which was taken as indicating the liquid-vapor spinodal (also with lines to guide the eye). The solid lines show the liquid-liquid coexistence line, which terminates at the LLCP (squares), and the dotted lines show the location of the Widom line. Both of these were determined from the two-state model fit to the simulation results. The stars show the loci of isothermal compressibility maxima, also determined from the two-state model.

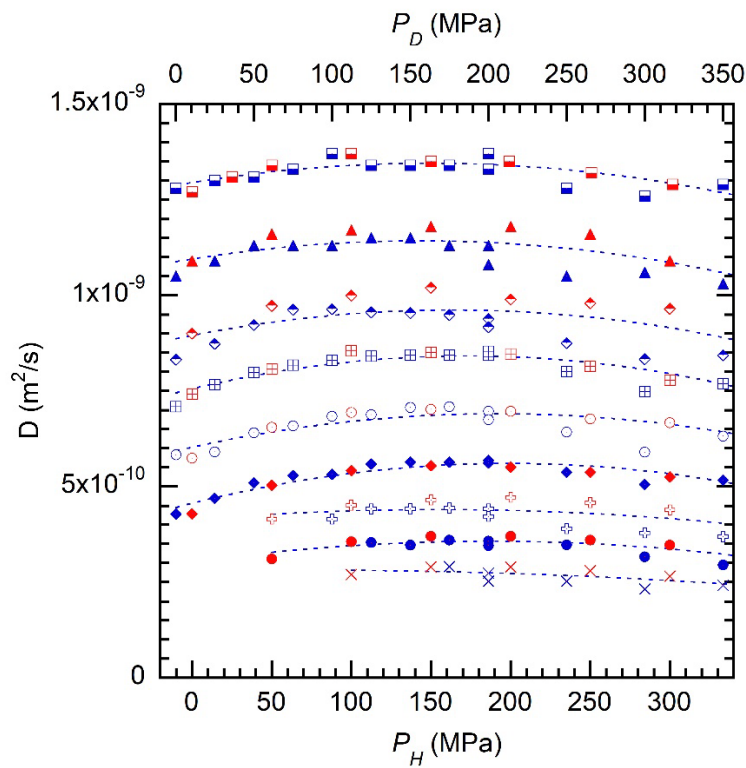

Fig. S8. Diffusion data for $H_2O$ (red symbols) and $D_2O$ (blue symbols) versus pressure.[5-7] The $D_2O$ data has been scaled by the square root of the masses. The temperatures, from top to bottom for the $H_2O$ ($D_2O$) data are 277.15 (283), 273 (278), 268 (272.5, 273), 263 (268), 258 (263), 252 (258), 248 (252.5, 253), 243 (248), 238 (243) K. Within the uncertainty of the diffusion data, it also appears to follow the low temperature correspondence found for the thermodynamics properties. Note that temperature pairs shown are not exactly corresponding according to Eqn. 2. However, other than the 268 (272.5) K pair, the differences between the experimental temperature pairs shown and corresponding temperatures are ≲ 0.6 K. The dotted lines are guides to aid viewing.